\documentclass[runningheads]{llncs}
\usepackage{algorithm}
\usepackage[noend]{algpseudocode}
\usepackage{graphicx}
\usepackage{amsmath}
\usepackage{graphicx}
\usepackage{amssymb}

\usepackage[english]{babel}
\usepackage[utf8]{inputenc}
\usepackage{algorithm}
\usepackage[noend]{algpseudocode}

\newcommand*{\field}[1]{\mathbb{#1}}%
\def\BibTeX{{\rm B\kern-.05em{\sc i\kern-.025em b}\kern-.08em T\kern-.1667em\lower.7ex\hbox{E}\kern-.125emX}}

\begin{document}
\title{Decentralized Cooperative Communication-less 
Multi-Agent Task Assignment with Monte-Carlo Tree Search }
%
\titlerunning{Abbreviated paper title}
%
\author{Mohammadreza Daneshvaramoli\inst{1} \and
Mohammad Sina Kiarostami\inst{1} \and
Saleh Khalaj Monfared\inst{1} \and
Helia Karisani\inst{1} \and
Hamed Khashehchi\inst{1} \and
Dara Rahmati\inst{1} \and
Saeid Gorgin\inst{2} \and
Amir Rahmati\inst{3}
}

\authorrunning{Daneshvaramoli et al.}
%
\institute{Institute for Research in Fundamental Sciences (IPM), Tehran, Iran  
\email{\{daneshvaramoli, skiarostami, monfared, h.karisani, hkhashehchi\}@ipm.ir}\\
\and
Iranian Research Organization for Science and Technology (IROST), Tehran, Iran
\email{gorgin@irost.ir}\\
\and
Stony Brook University, New York, USA\\
\email{amir@cs.stonybrook.edu}}
\maketitle              
\begin{abstract}
Cooperative task assignment is an important subject in multi-agent systems with a wide range of applications. These systems are usually designed with massive communication among the agents to minimize the error in pursuit of the general goal of the entire system. 
In this work, we propose a novel approach for Decentralized Cooperative Communication-less Multi-Agent Task Assignment (DCCMATA) employing Monte-Carlo Tree Search (MCTS). Here, each agent can assign the optimal task by itself for itself. We design the system to automatically maximize the success rate, achieving the collective goal effectively. 
To put it another way, the agents optimally compute each following step, only by knowing the current location of other agents, with no additional communication overhead. In contrast with the previously proposed methods which rely on the task assignment procedure for similar problems, we describe a method in which the agents move towards the collective goal. This may lead to scenarios where some agents not necessarily move towards the closest goal. However, the total efficiency (makespan) and effectiveness (success ratio) in these cases are significantly improved.
 To evaluate our approach, we have tested the algorithm with a wide range of parameters(agents, size, goal). Our implementation completely solves (Success Rate=\%100) a $20\times20$ grid with 20 goals by 20 agents in 7.9 s runtime for each agent. Also, the proposed algorithm runs with the complexity of
 $\mathcal{O}(N^2I^2 + IN^4)$, where the I and N are the MCTS iterative index and grid size, respectively.

\end{abstract}

\section{Introduction}

The emergence of the agent-based systems has drawn a tremendous attraction in computer science and software systems in recent years like the importance of parallelization in hardware approaches for problem-solving \cite{monfaredgenerating}.  This attraction is generally due to the fact that the multi-agent systems (MAS) are deemed to be remarkably effective in solving large-scale and sophisticated problems \cite{ferber1999multi}. MAS has recently hiked in popularity to solve a wide range of applications. It has been employed in a variety of complex and sophisticated tasks for its efficient, effective, and accurate performance. By reducing the execution time significantly, multi-agent systems are used widely in the robotic industry \cite{ota2006multi,arai2002advances} as well as in computer games \cite{adobbati2001gamebots,wang2008fast,kiarostami2019multi}, geographic information system \cite{lombardo2004intelligent} and many other applications like online trading \cite{luo2002multi} and social structure modeling \cite{sun2004simulating}.
\par By dividing the final goal of the entire system into multiple related subtasks, a multi-agent system improves the time complexity, while maintaining the accuracy at an acceptable level. There are different aspects of designing a MAS. Generally a MAS could be either employed \textit{centralized} or \textit{decentralized}. In a \textit{centralized} fashion where a central unit is responsible for deciding the collective state of the entire system and all of the agents follow the global policy computed by the central unit\cite{sharma2016comparative}. In such systems, usually data is communicated between each agent and the central unit, and the policy for each unit is transmitted by the central unit to each agent. However, in \textit{decentralized} and \textit{distributed} MAS which are more favoured in the literature nowadays \cite{xuan2002multi,ota2006multi,gan2011multi}, the collective goal is achieved by collective behaviour. This means that the agents communicate with each other and based on the collected data from the neighbour agents. In this way, the final decision is made by each agent for itself. This scheme, of course, requires a level of collective intelligence for the agents and a common policy which is shared between the agents. Moreover, the cooperation between the agents is inevitable in this design. This cooperative execution is prominent due to the fact that the main goal of the design is only feasible if the agents operate at each state based on the collected data from each other\cite{guttman1998cooperative,chaimowicz2002dynamic,koenig2017case}. 
\par Task assignment in a MAS is one of the challenging issues to optimally balance out the time complexity and success rate (accuracy) of the system. Many substantial efforts have been made recently in the literature to tackle this issue and find an efficient approach in the context of the problems \cite{khamis2015multi,lerman2006analysis,tang2007complete}. However, due to the inherent complexity of the \textit{Multi Agent Task Assignment} (MATA) algorithms in these works, the memory and computation consumption of these approaches are non-trivial and usually yields in some level of limitations in the system's overall performance. In other words, one of the most significant aspects of designing a MAS is to efficiently assign the tasks in a manner to maximize the performance and accuracy while keeping the collision and time-complexity minimized. Moreover, proposing a general perfect solution to assign tasks to multiple agents is known to be crucially hard since the complexity and accuracy of the solution is extremely dependable on the parameters of the problem/puzzle itself. 
\par In this work, we propose an extremely efficient method to tackle this issue by proposing a progressive methodology based on Monte-Carlo Tree Search (MCTS). The term \textit{progressive} approach here, refers to a progressive probabilistic approach where instead of initial task assignment, agents are progressively moved to their optimal goals in a collective behavioral action saving a tremendous computational-memory consumption, which also leads to minimizing the collision between the agents. The MCTS has shown promising results in the same category of the problems \cite{6145622}. In this work, we propose a modified \textit{Multi Agent Monte-Carlo Tree Search} (MAMCTS) with some specific optimization to optimally solve the problem in hand. Moreover, we show that the proposed method is effectively implementable when the scale of the problem increases.
\par The contributions of this work are:
\begin{enumerate}
  \item We have proposed a probabilistic method based on Monte-Carlo Tree Search to approach the defined problem. The proposed method successfully execute the task assignment operation with 100\% success rate in polynomial order of $\mathcal{O}(N^2I^2 + IN^4)$.
  \item The mathematical foundation of the problem is described, and a decentralized solution with minimum communication among the agents is proposed.
  \item We have proposed a couple of optimizations in the Monte-Carlo Tree Search, specified for the problem which reduces complexity in the simulation process.
\end{enumerate}

\par The remaining of the paper is organized as follows:
Section 2 gives a study regarding the background of the problem, offering some mathematically formulated definition of the problem. Section 3 investigates the related efforts in the field during recent years. We thoroughly discuss our methodology in Section 4 with the implemented algorithms and the proposed optimizations. In section 5, we evaluate our proposed work by employing some standard benchmarks in time and accuracy criteria. Finally, section 6 concludes the paper and gives some discussions regarding the current and future works.


\section{Background}
In this section, first, we would like to comprehensively discuss Monte-Carlo Tree Search (MCTS), focusing in its algorithm in detail and then, we will precisely explain Multi-Agent Task Assignment or Allocation (MATA) problem with a profound mathematical approach. Finally, a real-world application of MATA will be hashed out to show the importance of this subject. 

\subsection{Monte-Carlo Tree Search (MCTS)}
Generally, MCTS is a best-first search, which employs Monte Carlo methods to sample steps or operations in a problematical manner in a specific domain \cite{6145622}. MCTS is one of the best approaches for solving Tree-based problems because of providing a guarantee for finding a reliable optimal or near-optimal solution and also its low time-complexity. This algorithm has four main steps which are called \textit{Selection}, \textit{Expansion}, \textit{Simulation}, and \textit{Backpropagation} which is fully discussed in \cite{6145622}. First of all, the MCTS algorithm, like other tree-based approaches, constructs a tree from a root which in this context, the root is the beginning cell of the agent. In the selection phase, the algorithm steers the search tree based on the Tree Policy to a leaf node. In the expansion step, other non-terminal children of the current node if exist, will be expanded. Next step is the simulation. In this step, the algorithm starts to move forward randomly in the nodes of the tree based on a policy, until it accomplishes a result. Finally, in the backpropagation step, the results of the previous step are propagated up, and then the selection step of the next round will take effect from it. Simple MCTS algorithm steps are presented in Algorithm \ref{alg:1}.

\begin{algorithm}
\caption{Monte-Carlo Tree Search}\label{alg:1}
\begin{algorithmic}
\Procedure{MCTS}{$root$}
\While { within~computational~budget}

    \State leaf = selection($root$)
    \State expandedLeaf = expansion($leaf$)
    \State simulationResult = rollout($expandedLeaf$)
    \State backpropagation($simulationResult, expandedLeaf$)
\EndWhile\textbf{end while}
\State \textbf{return} bestChild($root$)
\EndProcedure
\end{algorithmic}
\end{algorithm}

\subsection{Multi-Agent Task Assignment (MATA)}

The term MATA is generally referred to as assigning multiple tasks to execute agents. However, in this work, we use an entirely different approach which relies on assigning \textit{Path} to agents. In this section, we give a mathematically formulated definition of the Multi-Agent Path Assignment(MAPA) problem to precisely investigate the effect of altering the underlying parameters to optimize the desired criteria which will be thoroughly discussed.
\par Consider the set of agents $A$ where $N_A=|A|$ represents the number of the agents. Also, the set of goals $G$ where $N_G=|G|$ in an
${N}\times {N}$ 
grid is considered. The function $\pi$ identifies the movement path in the grid and is defined as follows:\\
\begin{equation}  \label{eq1}
\begin{split}
    &\pi : \mathcal{\field{N}}\times \mathcal{\field{N}}\rightarrow\mathcal{\field{N}}\times \mathcal{\field{N}}
\end{split}
\end{equation}
Here, we would like to define an efficient $\pi$ to maximize some criteria. Firstly, for any agents $i\in A$, $\pi(i,t)$ demonstrates the location of the agent number $i$ at time $t$ in a discrete time-line representation. Some conditions are defined as follows to describe the MAPA problem.

\begin{equation} \label{eq:2}
\begin{split}
 &\pi(i,0)=s_i\\
 &\forall i\leq N_A,t \in \field{N}:  MD(\pi(i,t), \pi(i,t+1)) \leq 1\\
 &\forall i,j \in A \quad:\quad \pi(i,t) \neq \pi(j,t)\\
\end{split}
\end{equation}
 
 In Equation \ref{eq:2}, each agent $i \in A$ is initially positioned at source $s_i$. The $MD()$ refers to the \textit{Manhattan Distance}, which defines the neighbour cells in terms of discrete movement. This means that each agent either moves to the neighbour cell or stays still in the same cell. Moreover, for each agent arriving at a goal, the agent stops. This is presented as below:
 
\begin{equation} 
\begin{split}\label{eq:3}
 &\forall \pi(i,t) \in G \Rightarrow \pi(i,t) = \pi(i,t+1)
\end{split}
 \end{equation}
 The main problem in the defined problem here is to maximize the number of agents arrived at the goals in some specified time constraint ($t_{final}$) as defined below. 
\begin{equation} 
\begin{split}\label{eq:4}
 &Max|(\bigcup_{i\in A}\{\pi(i,t_{final})\})\bigcap G|
\end{split}
 \end{equation}
 
 Note, that $t_{final}$ could be limited by the policies of the problem in accordance with the application constraint. This will definitely impact the accuracy of the results since the time steps for agents are now limited. Different constraint and criteria for the problem could be handled more efficiently by some modifications in implementation. These points and the proposed solutions for dynamic \textit{$t_{final}$} will be addressed and discussed.
\section{Related Efforts}
The described problem here is very similar to the defined scenarios by other works, which will be briefly discussed in this section.
The Multi-Agent Path Finding problem is extensively employed to be utilized in large-scale, sophisticated problems. 
 The standard MCTS is used to handle simple cooperative problems \cite{7332728}. Kartal et al. \cite{Kartal:2014:UNV:2615731.2615746} presented a framework for generating believable narratives by using MCTS and also proposed a centralized approach with MCTS to solve the multi-robot task allocation problem with time and capacity constraints \cite{kartal2016monte}. Likewise, Lenz et al. \cite{7535424} contributed a cooperative motion planning algorithm based on MCTS without communication. Godoy et al. \cite{10.1007/978-3-642-33932-5_56} showed the trade-off between space, time, and communication affects the number of tasks completed. Pierson et al. \cite{7801073} proposed a decentralized algorithm for the cooperative pursuit of multiple evaders.

\par Kiarostami et al. in \cite{kiarostami2019multi} introduced a multi-agent non-overlapping pathfinding with an enhanced MCTS by reducing \textit{branching factor} of the search tree. Gelly and Silver \cite{gelly2011monte} introduced two major enhancements to Monte-Carlo tree search and then applied them to \textit{Computer Go}. The first one represents a faster estimate of the action value in comparison to the normal MCTS and the second enhancement is using a heuristic function learned by temporal-difference learning and self-play to initialize the values of new positions in the search tree.
\\ The Multi-Agent or Multi-Robot task allocation problem is also, another major area which has been investigated by many researchers. Chaimowicz et al. \cite{chaimowicz2002dynamic} represented a role assignment under a hybrid system framework, using a hybrid automaton which leads to having an adaptive system to unexpected events and having efficient performance.
\\ As another instance, Shaheen Fatima and Wooldridge \cite{fatima2001adaptive} addressed the problem of assigning tasks between agents by themselves efficiently and proposed an adaptive organizational method for time-constrained domains with varying the computational load.\\
\par Zerbal et al. \cite{phdthesis} proposed a solution for the exact defined problem which fully resolves the small-size scenarios and partially solves the bigger problems. However it does not discuss the run-time and efficiency of the their solution.
\par Despite of the similarities between the defined problems in the related literature, all of the cited previously proposed method, does not guarantee the full solution of the problem. However, we have proposed a method for a similar problem which successfully solves the different configurations of the problem with full accuracy.
\section{Proposed Method}
In this section, our novel solution for DCCMATA problem has been thoroughly discussed. Firstly, some mathematical preliminaries are outlined and defined. Then, we propose the MA-MCTS algorithm to solve the problem with respect to time and accuracy criteria, respectively. 
\subsection{Preliminaries}
\begin{figure}[H]
  \includegraphics[width=0.9\textwidth]{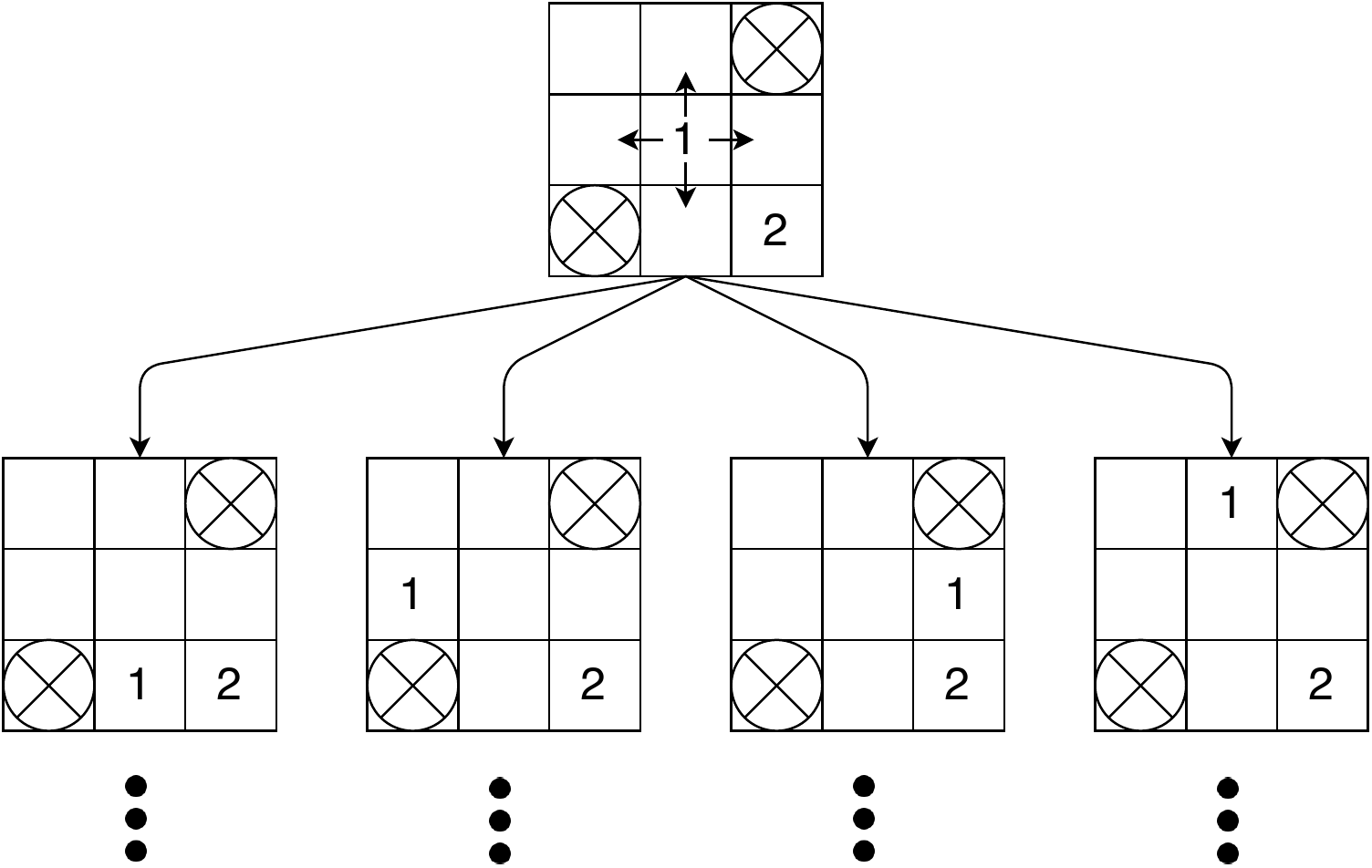}
  \centering
\caption{Parent and Children relationship in MCTS} 
\label{tree}
\end{figure}

Before getting into the details it is important to notify that the agents are moving simultaneously in the grid and the following clarifications and definitions are addressed here for the sake of simplicity and explanation for each agent.\\
Each node in the Monte-Carlo Tree Search is defined by a $N\times N$ table which contains the current positions of all of the agents. We refer each of these nodes as the problem \textit{State} based on the values stored in its table. We should note that the children of a node differ from their parents in a single entry of the table. This relation between parent and children is based on the Equations stated \ref{eq:2}, which is also demonstrated in Figure \ref{tree}.

\subsubsection{Makespan}
Each step of the MCTS here is a single \textit{Time Step} which will be referred to \textit{makespan} from now on. By incorporating this definition, one can enforce a time limitation to the problem which will be named by \textit{$T_{final}$}. It is clear that by this limitation, the maximum depth of the MCTS will be set to \textit{$T_{final}$}, which basically means the maximum permitted makespan to solve the problem. consequently this leads to maximum \textit{$T_{final}$} moves for each agent.
\subsubsection{Accuracy}
The final accuracy of the solution could be easily regarded as the success ratio of the solution. This is simply illustrated in Equation \ref{eqacc}.

\begin{equation} 
\begin{split}\label{eqacc}
 &Success Rate=\frac{N_{Achieved Goals}}{N_A}
\end{split}
 \end{equation}
\subsubsection{Time-Accuracy}
As discussed earlier, in the case of limitation in \textit{$T_{final}$}, the optimal solution is to maximize the function addressed in Equation \ref{eq:4}. As, MCTS approach the problem in a probabilistic manner, we argue regarding this trade-off in the evaluation section. 
\subsection{Multi-Agent MCTS}

The MCTS is thoroughly discussed in section. The proposed method employs a Multi-Agent MCTS (MA-MCTS) implementation. The MA-MCTS is a variation of MCTS which is simply executed on all of the agents according to their environmental input. Algorithm \ref{alg:2} demonstrates the high level of the MA-MCTS.

    
    

\begin{algorithm}
\caption{Multi-Agent Monte-Carlo Tree Search}\label{alg:2}

\begin{algorithmic}
\Procedure{MAMCTS}{$grid,T_{final},agents$}
\While{$T_{cur} < T_{final}$ and Problem Not Resolved}
\For {each Agent} 
\For {MCTS Iteration} 
    \State Selection
    \State Expand Tree if Node Is Not Terminal 
    \State Rollout Using Uniform Random Policy
    \State Back-Propagation
    \EndFor\textbf{end for}

    \State Select Best Policy
    \EndFor\textbf{end for}
    
    \For { each Agent}
    
    \State Merge States
    \vspace*{1mm}
    \EndFor\textbf{end for}
    \For { each Agent}
    \State Copy State
    \vspace*{1mm}
    \EndFor\textbf{end for}
        \vspace*{1mm}
\EndWhile{\textbf{end while}}
\EndProcedure
\end{algorithmic}
\end{algorithm}

\begin{figure}[t]
  \includegraphics[width=0.9\textwidth]{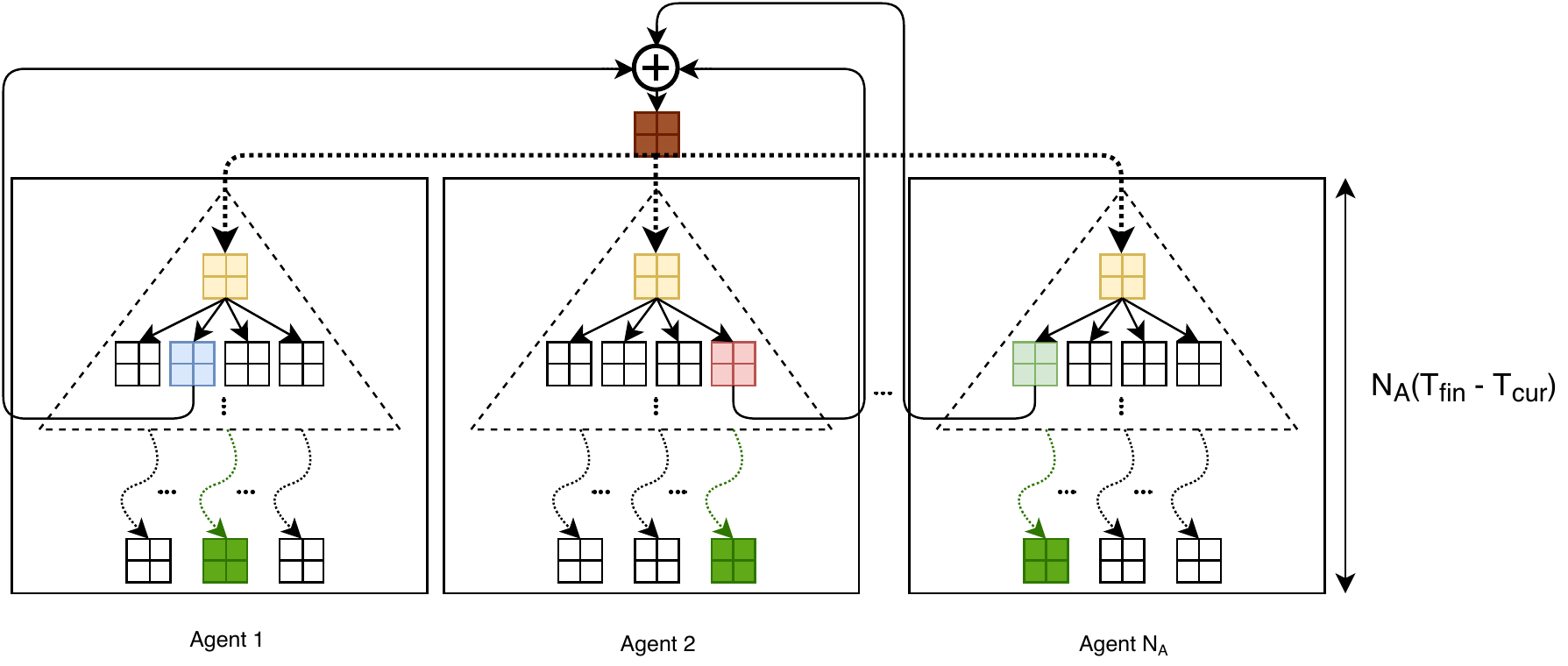}
  \centering
\caption{The structure of the proposed MA-MCTS } 
\label{mcta}
\end{figure}

According to Algorithm \ref{alg:2}, and in the case of our problem, the MA-MCTS should be performed for each agent in parallel. Hence, each agent executes the MCTS for every possible outcome of the grid from the current state up to the final state of the problem. Assuming $N_A$, as the number of the agents, the MCTS explores a tree with depth of $N_A \times (T_{final}-T_{Current})$.\\ 
Figure \ref{mcta} represents the overall structure of the proposed MA-MCTS. The deterministic exploration of the MCTS is performed for $M$ children (iterations), and the next states are randomly chosen at the simulation stage.
Furthermore, as illustrated in Figure \ref{tree} the \textit{Branching Factor} of the MCTS here is equal to 4 since agents are permitted to move a single step in the 2-D grid. By constricting the Monte-Carlo tree for each agent, the system is autonomously initiated. At each stage, after selecting the desired next step by evaluating the \textit{Reward Value}, each agent updates its position based on the chosen state. Then passing a single time-step, by combining the updated positions of the agents, a single unique state is constructed for all of the agents and it is fed into the MCTS as the new parent node. In this iterative procedure, as the $T_{current}$ increases, the depth of the MCTS decreases, so the problem gets easier toward the final state. \\
Taking into account the functions and the procedures in the proposed method, the complexity order of the algorithm could be computed as follow:
\begin{equation} \label{eq:12}
\begin{split}
    \mathcal{O}(N_AI^2T_{final} + N_AIT_{final}^2 + N_AT_{final}IN^2)=\mathcal{O}(N^2I^2 + IN^4)
\end{split}
\end{equation}

\subsubsection{Reward Function}
The \textit{Reward Function} in the MA-MCTS in the proposed algorithm is dynamically tuned in the algorithm structure. At each node of the MA-MCTS the value of each node could be naively computed by simply dividing number of captured goals by the number of agents at each state, as following in Equation \ref{eq:5}:\\
\begin{equation} \label{eq:5}
\begin{split}
    &Value_{naive}=\frac{N_{Goals Captured}}{N_A}
  \end{split}
\end{equation}

The main problem with the aforesaid formulation is that the cooperative
action of agents is not considered. The other important shortcoming is the fact
that the effect of finding a goal by the agent does not impact the value at all.
This major issue clearly, would also cause some serious difficulties regarding the
cooperative nature of the problem since agents have no clue weather their action
have led the system toward the collective goal or not. Keeping in mind that the
studied system here is completely communication-less and cooperative. We can
come up with a modified Value Function which considers the addressed issues
as follows:
\\

\begin{equation} \label{eq:6}
Value_{mod} = 
\begin{cases}
 \frac{N_{Goals Captured}}{N_A}-\frac{\alpha}{N_A}& \text{if} \textit{ Mark} \text{ = 1} \\
  \frac{N_{Goals Captured}}{N_A} & \textit{ Otherwise}
\end{cases}
\end{equation}

In Equation \ref{eq:6}, \textit{Mark} is Boolean value, which is set to 1 where the agent reaches a goal. Also, we define $\alpha$ to be a tuning variable where $\alpha \in \{0,1/2,1\}$. By employing \ref{eq:6} as the \textit{Value Function} the impact of cooperative action, linearly affect the final decisions. By increasing $\alpha$ the cooperation level increases linearly. Note that to implement the cooperative actions, finding goals by other agents is more valuable compared to the states where the agent finds a goal for itself. Straightforwardly, to emphasize the collective goal, each agent cares more about other agents than itself.
Finally, after calculating each value, the previous value should be updated. We have employed two different approaches to update the value which are demonstrated in Equations \ref{eq:7} and \ref{eq:8}.

\begin{equation} \label{eq:7}
\begin{split}
    &Value_{New}=\frac{(K-1)\times Value_{old}+Value_{mod}}{K}
  \end{split}
\end{equation}
\begin{equation} \label{eq:8}
\begin{split}
    &Value_{New}=Max(Value_{mod},Value_{old})
  \end{split}
\end{equation}

Our experimental evaluation shows that the approach in Equation \ref{eq:8} is more successful in some more sophisticated scenarios, where the approach in Equation \ref{eq:7} is much more faster and converges to the final solution in most of the cases.\\
By changing three values for $\alpha$ and switching the value update functions, the total number of 6 scenarios for each simulation is considered in our solution, which will be thoroughly investigated in the evaluation section.
\subsubsection{Depth}
\begin{figure}[t]
  \includegraphics[width=0.9\textwidth]{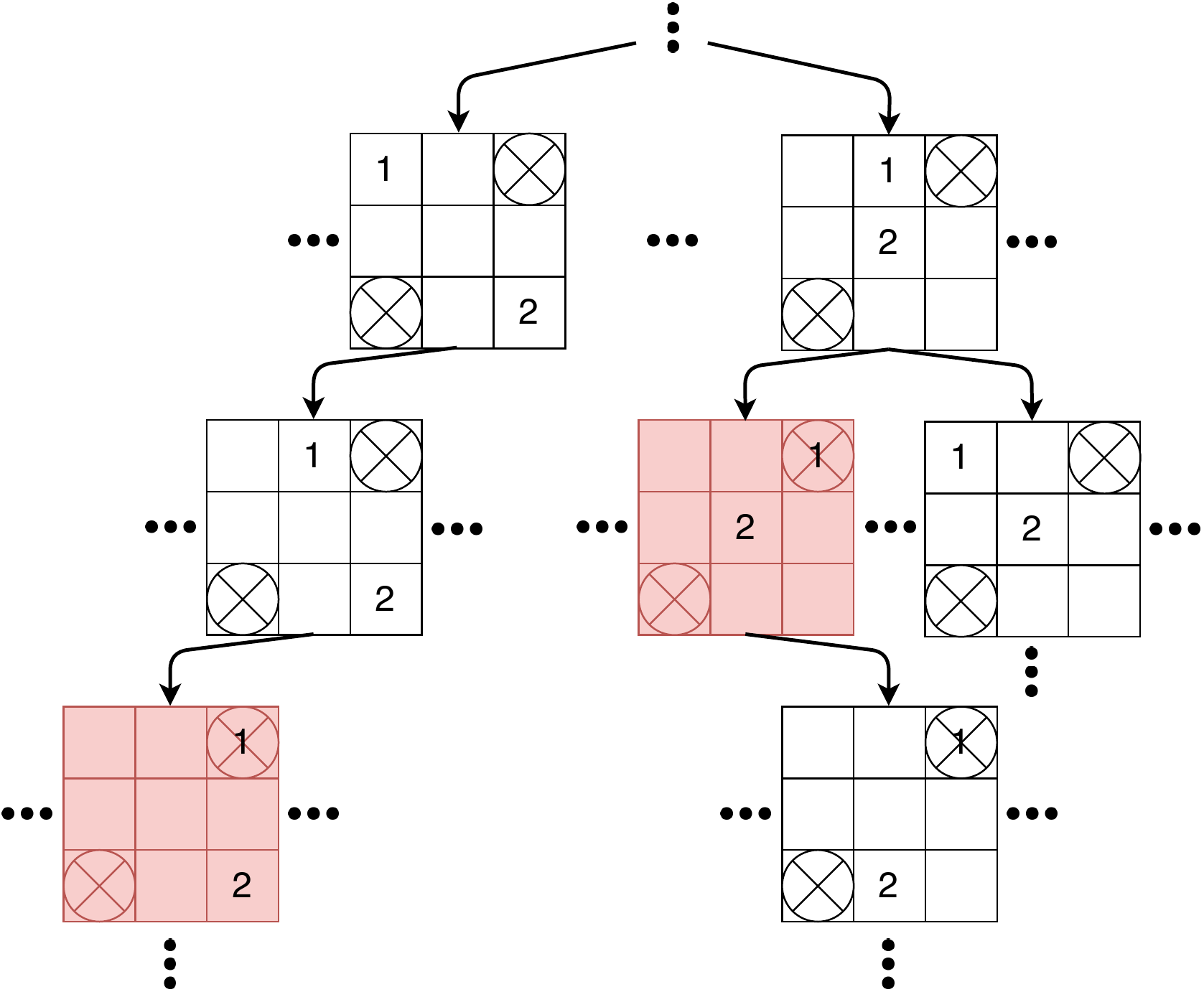}
  \centering
\caption{Preferences in the case of similar states leading to extension in depth.} 
\label{depth}
\end{figure}

The \textit{Depth} of the tree search, directly affects the accuracy and efficiency of the solution. Here, we have employed a dynamic approach to decrease the depth of the tree search with a trivial cost. The approach is to prefer nodes with lower height when nodes have similar values. This is accomplished by applying Equation \ref{eq:9} to the \textit{Value Function}.
\begin{equation} \label{eq:9}
\begin{split}
    &Value_{mod^{*}}=Value_{mod}+\frac{1}{N_A}\times(1-\frac{t_{current}}{t_{final}})
  \end{split}
\end{equation}
To demonstrate the simple modification governed by Equation \ref{eq:9}, Figure \ref{depth} shows the preference of the algorithm when similar values are returned.
In Figure \ref{depth}, the highlighted state (node) in the second layer at the right side, is preferred over the other node in the third layer, despite their similar state table. \\
Note that the proposed approach here decreases the search depth remarkably since the insignificant deeper nodes are more likely to be ignored.

\subsection{Optimization}
In this section, we address some of the optimizations in the proposed algorithm. Note that the optimizations discussed here could be applied to a wide range of different problems similarly. Also, it is worthy to note that some of the optimizations regarding the implementation, would highly improve the performance of the algorithm.
\subsubsection{Rollout Function}

The \textit{Rollout Function} stage of the MCTS algorithm involves random tree exploration. We have applied a novel technique in this function, which reduce the time complexity dramatically. Understandably, every state of the problem at each node is represented by a new grid with a size of $N \times N$.So, this computation for nodes causes a $\mathcal{O}(N^2)$ complexity for each node. Also, note that this computation overhead is accumulated for each stage. However, in the case of our problem, the next stage differs from the parent state only in a single cell. So, instead of constructing a new $N \times N$ grid for each state, one could simply modify a single cell to generate a new child. Also, in the \textit{Rollout} procedure storing the states are unnecessary since only the final states are important for the evaluation. This optimization reduces the steering cost to $\mathcal{O}(1)$. It is also clear that numerous new sates should be constructed at each stage, which all are constructed in $\mathcal{O}(1)$. \\
Moreover, this technique is applicable in scenarios where the child node differs from its ancestor insignificantly. In such applications, it is relatively easier to construct a new child by modifying the parent node. However, there are different applications where child nodes are constructed with more complex formulation, and the discussed optimization does not actually improve the performance of the search procedure in terms of memory and time complexity.

\subsubsection{Tree Search vs. Graph Search}

\begin{figure}[t]
  \includegraphics[width=0.9\textwidth]{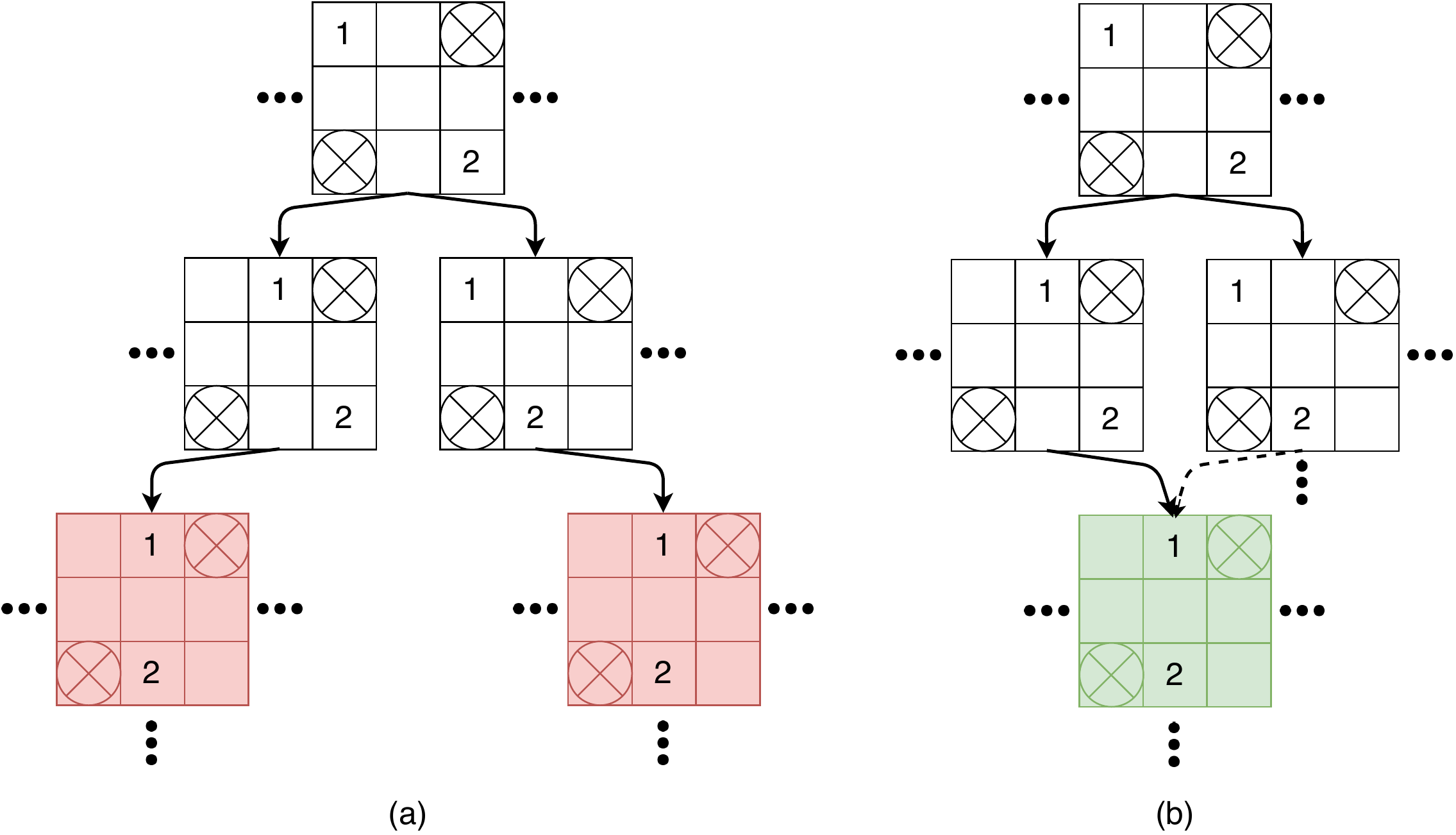}
  \centering
\caption{Differences between Tree Search (a) and Graph Search (b) approaches.} 
\label{opt1}
\end{figure}

There are generally two main approaches to execute any search procedure. The search algorithm is usually preformed by conventional \textit{Tree Search} which basically constructs a memory-less  structure to execute the search which results in a  fast tree exploration. On the other hand, \textit{Graph Search} keeps a \textit{closed} list of the possible branches for further time efficiency. Hence, employing graph search will significantly reduce the size of the tree for exploration. Figure \ref{opt1} highlights this difference. The \textit{Tree Search} approach showing Figure \ref{opt1} (a) will result in similar states regularly, while in Figure \ref{opt1} (b), by accessing memory based list similar states are explored only once.

In the case of our problem, the use of the \textit{Graph Search} certainly improves the efficiency of the tree exploration, since it will reduce the \textit{Branching Factor} and so the corresponding \textit{Depth} will significantly rise. 
Further in-depth performance profiling shows $\mathcal{O}(N^2)$ computational overhead regarding the construction of a \textit{Hash Table} to store the state of the nodes in the \textit{Graph Search} method. This construction and its validation are extremely expensive. So, we have employed a modified \textit{Tree Search} approach, which will be thoroughly discussed as follows.
\subsubsection{Tree Search: Variable ordering}
According to the nature of a communication less system, one should design a proper resolution to efficiently distribute the \textit{Tree Search} among the agents in order to maximize the overall performance. This is achieved by reordering the \textit{Search Tree} procedure according to the \textit{Agent Number} for each agent.
Similarly, to compensate the shortcoming addressed in the previous section regarding the \textit{Tree Search} procedure, we have Incorporated an optimization in the ordering and organization of the \textit{Tree search} itself. To avoid the searching the similar states in the \textit{Tree Search} as shown in Figure \ref{opt1} (a), instead of switching among agents at each stage, every child for a specific agent is constructed and is explored, followed by all of the children for the next agents and so on. In such a structure, similar states do not occur at least for $N_A+1$ levels in tree.
More precisely, by employing this optimization the possibility of occurring similar states is significantly reduced.

\section{Experimental Results and Discussion}

In this section, we implement our proposed algorithm on different scenarios and a wide range of configuration. We discuss our evaluated results based on different criteria and parameters to show the efficiency and scalability of the proposed approach.\\
We have evaluated our implementation based on time-limited and fully accurate scenarios which are extensively discussed.

\subsection{Setup}
 We have implemented our algorithm by using Java 12 platform. The compiled Jar file is executed on a machine running Ubuntu 16.04 equipped with two Intel XEON E5 2697V3 CPUs clocked at 2.6 GHz. Moreover, the system is supplied with 128 GB of DDR3 RAM.
\subsection{Full Accuracy Evaluation}
By randomly locating a different number of agents with the same numbers of workload positioned in uniformly random cells, a different number of instances are constructed.
\\
For each instance, tuning the $\alpha=\{0,1/2,1\}$ according to the Equation \ref{eq:7} and switching the \textit{Reward Value} in accordance to Equations \ref{eq:8} and \ref{eq:9}, leads to different results. To ensure that the solution reach the \textit{Success Rate} of \%100 (full accuracy) the $T_{final}$ is experimentally set to the large value of $3\times(N)$. However, the program will exit immediately after reaching the \%100 SR (When all the goals are captured by agents).
\\Table \ref{tab:1} represents the evaluation of the proposed algorithm with no time-limitation. \textit{Total Time} refers to the total elapsed time for all agents to successfully capture a goal if the implementation is sequential. However, In the real scenario implementation, the MCTS are executed separately in each agent. In this case, the \textit{Average Time} and the \textit{Maximum Time} elapse by the agents are shown to represent the actual real-world scenario. Note that, the MCTS iteration index in this implementation is set to $I=2000$, which means that 2000 numbers of children are selected and the search is executed. Also, to normalize the results in Table \ref{tab:1}, 50 iterations of each instance are simulated and all of the reported \textit{Run-Time} are averaged.

\subsection{Accuracy-Time Evaluation}
\begin{figure}[t]
  \includegraphics[width=0.9\textwidth]{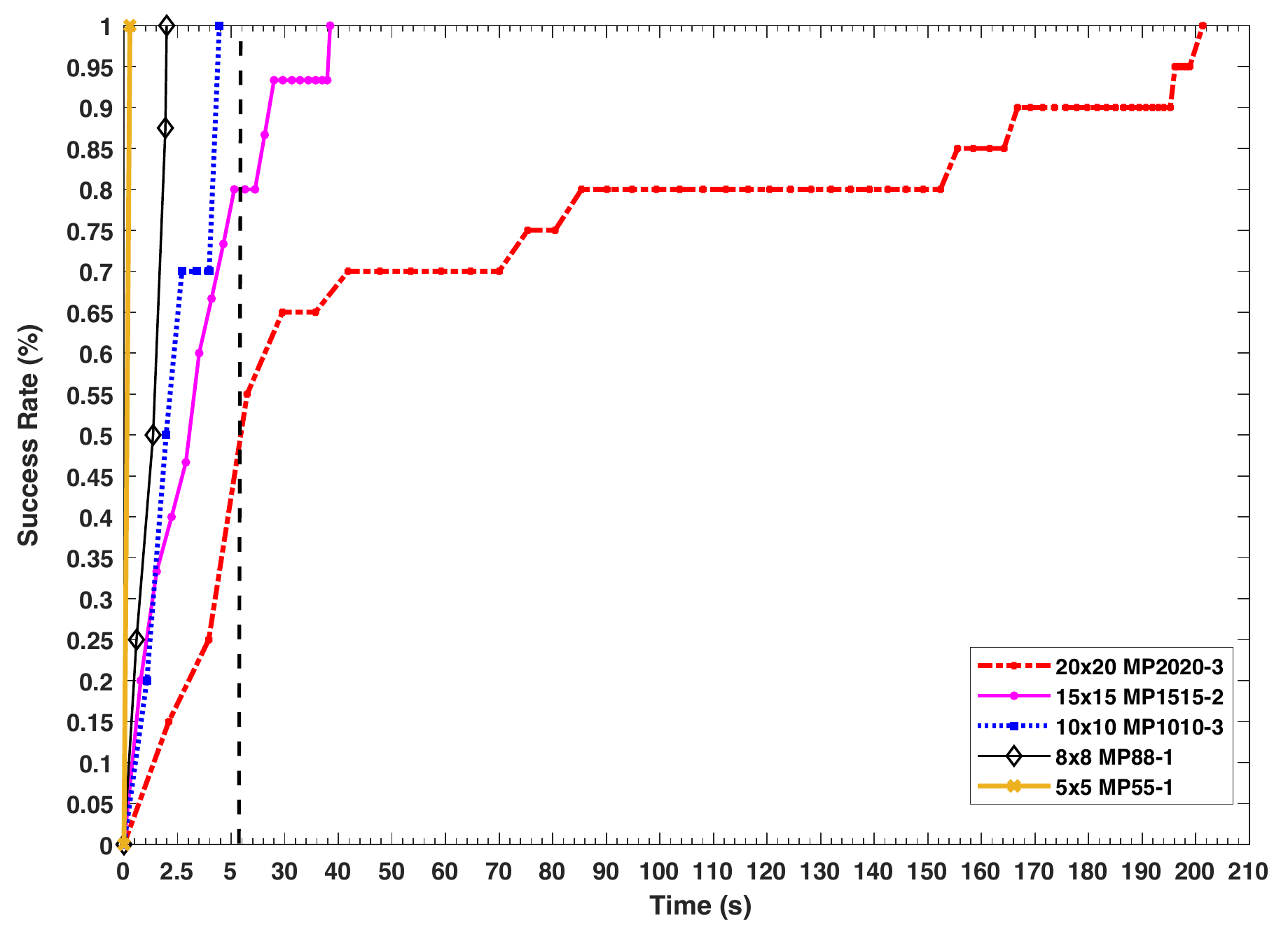}
  \centering
\caption{Time-Accuracy evaluation of the proposed method.} 
\label{gifres}
\end{figure}

When the time limitation is enforced by the user to solve the problem, the MCTS depth decreases, which results in a different tree exploration in the search procedure. Figure \ref{gifres} shows the Accuracy-Time trade-off for different instances.
\begin{table}[t] 
 \caption{Evaluation of solutions with no time restriction (Full-Accuracy)}
 \label{tab:1}
\begin{tabular}{|c|c|c|c|c|c|c|c|}
\hline
\begin{tabular}[c]{@{}c@{}}Grid \\ Size($N\times N$)\end{tabular} & \multicolumn{1}{l|}{Instance} & \begin{tabular}[c]{@{}c@{}}Agents \\ Num($N_A$)\end{tabular} & \begin{tabular}[c]{@{}c@{}}Full Search\\  Space($4^{3NN_A}$)\end{tabular} & \begin{tabular}[c]{@{}c@{}}Total \\ Time(s)\end{tabular} & \multicolumn{1}{l|}{\begin{tabular}[c]{@{}l@{}}Average \\  Time(s)\end{tabular}} & \multicolumn{1}{l|}{\begin{tabular}[c]{@{}l@{}}Maximum \\    Time(s)\end{tabular}} & \multicolumn{1}{l|}{\begin{tabular}[c]{@{}c@{}}Maximum\\ Num of Steps\end{tabular}} \\ 
\hline
{5}$ \times ${5} & MP52-1   & 2                                                    & $4^{30}$                                             & 0.020                                                    & 0.011 &0.011                                            & 3\\ \cline{2-8} 
                                                        & MP53-1                        & 3                                                        & $4^{45}$                                             & 0.041                                                    & 0.013                                                                            & 0.014                                                                              & 3                                                                                   \\ \cline{2-8} 
                                                        & MP55-1                        & 5                                                        & $4^{75} $                                            & 0.276                                                    & 0.055                                                                            & 0.062                                                                              & 5                                                                              \\ \hline
{8}$ \times ${8}                                      & MP82-1                        & 2                                                        & $4^{48}$                                  & 0.040                                                    & 0.021                                                                            & 0.021                                                                              & 4                                                                                   \\ \cline{2-8} 
                                                        & MP83-1                        & 3                                                        & $4^{72} $                                            & 0.291                                                    & 0.100                                                                            & 0.105                                                                              & 6                                                                                   \\ \cline{2-8} 
                                                        & MP84-2                        & 4                                                        & $4^{96} $                                            & 0.345                                                    & 0.091                                                                            & 0.106                                                                              & 3                                                                                   \\ \cline{2-8} 
                                                        & MP85-2                        & 5                                                        & $4^{120} $                                           & 0.609                                                    & 0.121                                                                            & 0.133                                                                              & 4                                                                                   \\ \cline{2-8} 
                                                        & MP88-1                        & 8                                                        & $4^{192}$                                            & 1.992                                                    & 0.249                                                                            & 0.336                                                                              & 6                                                                                   \\ \hline
{10}$ \times ${10}                                     & MP102-2                       & 2                                                        & $4^{60}      $                                       & 0.206                                                    & 0.096                                                                            & 0.096                                                                              & 10                                                                                  \\ \cline{2-8} 
                                                        & MP103-2                       & 3                                                        & $4^{90}$                                            & 0.255                                                    & 0.088                                                                            & 0.095                                                                              & 4                                                                                   \\ \cline{2-8} 
                                                        & MP105-3                       & 5                                                        & $4^{150} $                                           & 2.110                                                    & 0.436                                                                            & 0.471                                                                              & 10                                                                                  \\ \cline{2-8} 
                                                        & MP1010-3                      & 10                                                       & $4^{300} $                                           & 4.450                                                    & 0.447                                                                            & 0.558                                                                              & 6                                                                                   \\ \hline
$ {15}\times ${15}                                     & MP152-3                       & 2                                                        & $4^{90}        $                                     & 0.202                                                    & 0.101                                                                            & 0.101                                                                              & 7                                                                                   \\ \cline{2-8} 
                                                        & MP153-1                       & 3                                                        & $4^{135}$                                            & 1.623                                                    & 0.558                                                                            & 0.593                                                                              & 13                                                                                  \\ \cline{2-8} 
                                                        & MP155-1                       & 5                                                        & $4^{225}$                                            & 5.443                                                    & 1.107                                                                            & 1.139                                                                              & 23                                                                                  \\ \cline{2-8} 
                                                        & MP157-1                       & 7                                                        & $4^{315}$                                            & 7.034                                                    & 1.016                                                                            & 1.121                                                                              & 12                                                                                  \\ \cline{2-8} 
                                                        & MP1515-2                      & 15                                                       & $4^{675}$                                            & 38.477                                                   & 2.569                                                                            & 2.623                                                                              & 20                                                                                  \\ \hline
{20}$ \times ${20}                                     & MP202-1                       & 2                                                        & $4^{120} $                                          & 0.683                                                    & 0.346                                                                            & 0.346                                                                              & 12                                                                                  \\ \cline{2-8} 
                                                        & MP203-1                       & 3                                                        & $4^{180}  $                                          & 1.777                                                    & 0.582                                                                            & 0.592                                                                              & 14                                                                                  \\ \cline{2-8} 
                                                        & MP205-2                       & 5                                                        & $4^{300} $                                           & 7.269                                                    & 1.436                                                                            & 1.501                                                                              & 20                                                                                  \\ \cline{2-8} 
                                                        & MP2010-1                      & 10                                                       & $4^{600}$                                           & 31.951                                                   & 3.197                                                                            & 3.364                                                                              & 21                                                                                  \\ \cline{2-8} 
                                                        & MP2020-3                      & 20                                                       & $4^{1200} $                                          & 201.325                                                  & 9.955                                                                            & 10.170                                                                             & 60                                                                                  \\ \hline
\end{tabular}
\end{table}

It is also interesting to notify that by tuning \textit{$T_{final}$}, the MCTS tends to find the optimal final state based on the enforced \textit{$T_{final}$}.

\section{Conclusion and Discussion}

With the emergence of sophisticated and complex problems and puzzles, Multi-Agent based solutions have been widely hired for their efficacy and performance. In this work, we have proposed a Multi-Agent Monte-Carlo Tree Search method suitable for large-scale communication-less task assignment problems.
\par By defining the problem with exact mathematical formulation, we strive to present an efficient and effective solution for the problem. We have designed a novel MCTS-based algorithm which builds a cooperative mechanism among the agents to achieve the collective goal. Furthermore, numerous optimizations for the proposed method are implied which improve the performance of the algorithm significantly. We have also discussed how these optimizations could be applied to other applications.
\par Considering different criteria and parameters, we have evaluated our work based on run-time and accuracy. In contrast with the recent works like \cite{phdthesis}, the proposed method here guarantees the successful solution on different scenarios achieving \%100 accuracy on a randomly generated multi agent grid while in \cite{phdthesis}, even a $5 \times 5$ grid had many fails which occured during its assignment in a random MAG. Furthermore, in \cite{Zerbel2019}, 1000 agents in a $100 \times 100$ grid had the same situation and many of the agents were failed. Moreover, our evaluation shows an acceptable run-time when a time limitation is enforced by the user.
\par For further improvement and study, one can implement the MCTS with a localized tree search approach which could reduce the run time dramatically. However, some consideration should be taken into account regarding the accuracy. Also, the impact of the different random distributions could be investigated.

%
%
%
\bibliographystyle{IEEEtran}
\bibliography{ms}

\end{document}